\documentclass{article}

\usepackage[position,nonatbib,final]{neurips_2026}
\usepackage{tcolorbox}
\tcbuselibrary{skins}

\usepackage{amssymb}
\usepackage{wrapfig}

\usepackage[utf8]{inputenc} % allow utf-8 input
\usepackage[T1]{fontenc}    % use 8-bit T1 fonts
\usepackage{hyperref,wrapfig}       % hyperlinks
\usepackage{url}            % simple URL typesetting
\usepackage{booktabs}       % professional-quality tables
\usepackage{amsfonts}       % blackboard math symbols
\usepackage{amsmath}
\usepackage{nicefrac}       % compact symbols for 1/2, etc.
\usepackage{microtype}      % microtypography
\usepackage{xcolor}         % colors
\usepackage{enumitem}

\definecolor{bgcolor}{RGB}{27, 71, 117}    % deep navy   — BGs
\definecolor{becolor}{RGB}{176, 87, 38}    % warm rust   — BEs
\definecolor{itocolor}{RGB}{38, 117, 117}  % teal        — ITO
\definecolor{glosscolor}{RGB}{38, 117, 117}  

\newcommand{\BG}{\textcolor{bgcolor}{$\blacklozenge$}}
\newcommand{\BE}{\textcolor{becolor}{$\bullet$}}
\newcommand{\ITO}{\textcolor{itocolor}{$\blacktriangle$}}

\newtcolorbox{methodcard}[2]{%
  enhanced,
  colback=#1!4!white,
  colframe=#1,
  coltitle=white,
  colbacktitle=#1,
  fonttitle=\bfseries\sffamily,
  title={#2},
  arc=2pt, boxrule=0.7pt,
  left=6pt, right=6pt, top=8pt, bottom=5pt,
  boxsep=2pt,
  attach boxed title to top left={xshift=8pt, yshift=-\tcboxedtitleheight/2},
  boxed title style={arc=2pt, boxrule=0pt},
}

\newcommand{\goal}[2]{%
  \par\smallskip
  \noindent\begin{tcolorbox}[
    enhanced, sharp corners, boxrule=0pt,
    colback=#1!12!white, leftrule=3pt, colframe=#1,
    left=4pt, right=4pt, top=2pt, bottom=2pt,
  ]
  \textcolor{#1!85!black}{\textbf{Goal.}}\ #2
  \end{tcolorbox}%
}

\newcommand{\term}[1]{\textcolor{glosscolor!75!black}{\textbf{\sffamily #1}}}

\newcommand{\glosscat}[1]{%
  \par\addvspace{2pt}%
  {\footnotesize\sffamily\bfseries\textcolor{glosscolor!65!black}%
    {\MakeUppercase{#1}}}\par
  \vspace{0.5pt}%
  {\color{glosscolor!25!white}\hrule height 0.4pt}%
  \par\vspace{0pt}%
}

\definecolor{checkcolor}{RGB}{27, 71, 117}    % navy, matches BGs
\definecolor{checkaccent}{RGB}{38, 117, 117}  % teal, matches glossary

\newcommand{\checksection}[1]{%
  \par\addvspace{8pt}%
  {\footnotesize\sffamily\bfseries\textcolor{checkaccent!65!black}%
    {\MakeUppercase{#1}}}\par
  \vspace{1pt}%
  {\color{checkaccent!25!white}\hrule height 0.4pt}%
  \par\vspace{6pt}%
}

\newlist{checklist}{itemize}{1}
\setlist[checklist]{
  label={\textcolor{checkcolor}{$\square$}},
  leftmargin=1.6em, labelsep=0.6em,
  itemsep=7pt, parsep=2pt, topsep=0pt,
}

\newcommand{\chkitem}[2]{%
  \textcolor{checkcolor}{\textbf{#1.}}\ \textit{#2}\par
  \vspace{1pt}%
}

\usepackage[backend=biber,style=numeric-comp,sorting=none,url=false]{biblatex}
\title{Looking Under the Streetlight:\\Evaluation in Generative Molecular Dynamics}

\author{%
 Simon Olsson\thanks{Corresponding author.} \\
  Department of Computer Science and Engineering\\
  Chalmers University of Technology and University of Gothenburg\\
  SE-41296 Gothenburg, Sweden\\
  \texttt{simonols@chalmers.se} \\
  \And
  Frank No\'e \\
  Microsoft Research Berlin\\
   \AND
  Grant M. Rotskoff \\
   Stanford University \\
   \And
   Kresten Lindorff-Larsen \\
   University of Copenhagen \\
}

\begin{document}

\maketitle
\vspace{-0.8cm}
% -----------------------------------------------------------------------
\begin{abstract}
This position paper argues that the generative molecular dynamics (GenMD) community is not properly evaluating its progress: current evaluations let a model match what is reported while being arbitrarily wrong on what was claimed, which risks over-claiming advantage over classical baselines. This reflects a structural misalignment between common evaluation practices and the scientific tasks the field sets for itself. The reason is that the ground truth---exhaustive molecular dynamics on the systems of interest---is computationally costly, and this cost is what motivates the field. With ground truth inaccessible, evaluation drifts toward convenient proxies, and the resulting norms entrench easily. We identify misaligned evaluation patterns that persist as a community-level equilibrium, and propose concrete observables, reporting standards, and benchmark designs, both computational and experimental, that better reflect what GenMD aspires to deliver. Our aim is to give the field shared language and tools for evaluation that match its scientific ambitions before today's conventions calcify into norms.
\end{abstract}

\section{Introduction}

Molecular dynamics (MD) describes the stochastic time evolution of molecules on a high-dimensional potential energy landscape. One can picture a ball rolling on a landscape of deep valleys and tall ridges, buffeted by random kicks from thermal noise. \begin{wrapfigure}{r}{0.55\linewidth}   \vspace{-0.4cm} 
\includegraphics[width=1.0\linewidth]{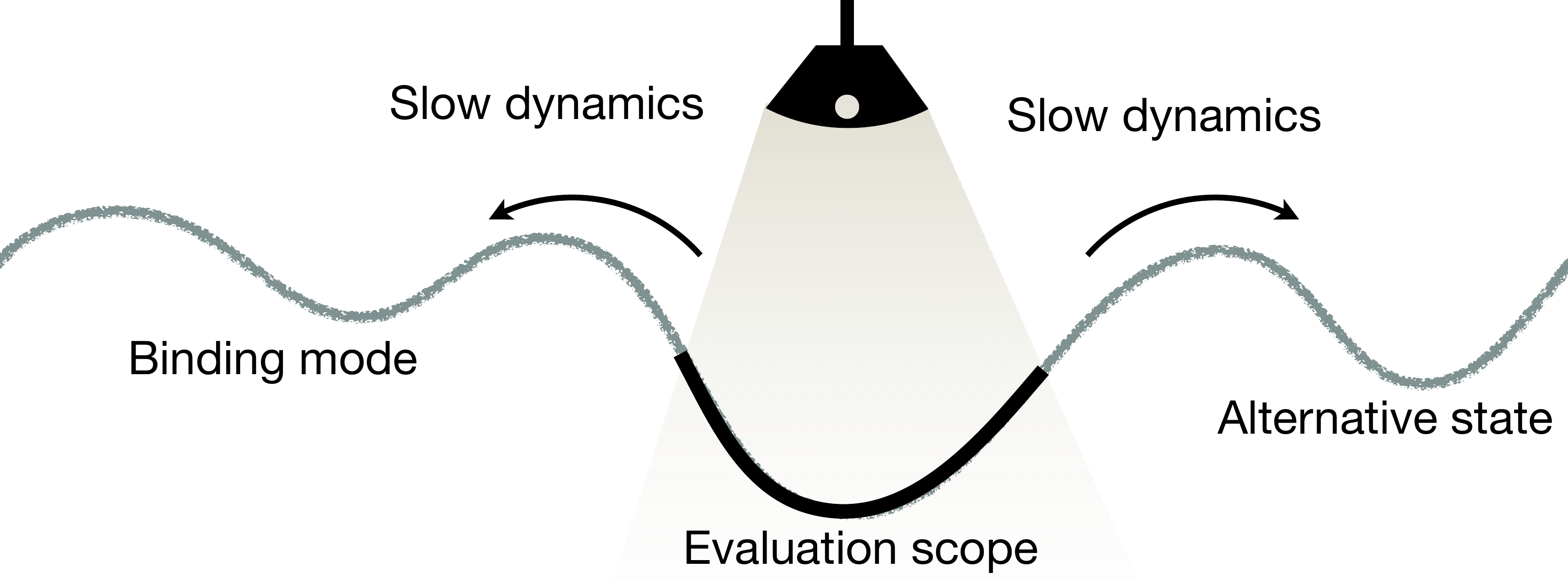}
    \caption{{\bf The evaluation streetlight effect:} short-trajectory references illuminate one basin, the evaluation scope; the rare events and slow modes that lead to alternative states and binding modes stay dark (P1, P2). }
    \label{fig:Illustration}\vspace{-0.2cm}
\end{wrapfigure}The statistics of motion within valleys, and of the rare crossings over ridges, correspond to high-dimensional experimental signatures measurable in the lab: MD provides a computational proxy for observable physical phenomena. This makes it a central tool across the sciences and engineering, from drug design to materials discovery, allowing properties of physical systems to be predicted before synthesis. Computations that reach experimentally relevant timescales are extremely expensive, and this cost sharply limits their practical impact.

Generative molecular dynamics (GenMD) has emerged as a promising research direction at the intersection of machine learning and computational physics. The central promise is significant: replace or augment expensive MD simulations with learned statistical surrogates that reproduce the same thermodynamic and kinetic quantities at a fraction of the cost~\cite{genmd}. Stated at its most ambitious, the goal is a {\it transferable} surrogate: a model that, given a molecular description (an amino acid sequence, a ligand--receptor pair, a materials composition), directly produces samples from the relevant thermodynamic or kinetic distribution without per-system simulation. Force fields such as AMBER~\cite{cornell1995amber} or CHARMM~\cite{mackerell1998charmm} are parameterized once and applied to any molecule within their scope; the analogous GenMD goal is the same generality at the sampling level.%, but per-system methods that require retraining for every new molecule do not achieve this: their cost scales with the number of systems studied, and in the limit they inherit the bottleneck they were meant to replace.

However, progress in GenMD is, by construction, measured against a baseline that is costly for the types of systems the field aspires to model. When the natural reference is inaccessible, evaluation drifts toward proxies that are easy to compute but largely insensitive to the slow, rare-event behavior that motivated the field. This is the streetlight effect at the community level (Fig.~\ref{fig:Illustration}): the only place lit by tractable reference data is rarely where the keys are.

\paragraph{Our position.} \textbf{The GenMD community is currently not properly evaluating its progress towards the goal it sets for itself and consequently risks systematically over-claiming its performance compared to classical baselines.} The amount of work resting on evaluations that cannot support their claims is growing, because the field's output is growing quickly while inherited evaluation conventions are not being re-examined; we do not claim that the fraction of misaligned evaluations is rising. The community has an opportunity to set standards for evaluating contributions at review time. Here, our aim is to state the goals of GenMD clearly, name the recurring patterns of evaluation that drift away from them, and propose practical remedies (observables, reporting standards, and benchmark designs) that the community can adopt. 

The patterns we identify: short reference data (P1), slow-mode-blind observables (P2), system selection narrower than the claims (P3), avoidance of self-correcting metrics (P4), absent classical baselines (P5), and cost accounting that ignores how the method will be used (P6). These share a common structure: each lets a model match what is reported while being arbitrarily wrong on what was claimed. Two demonstrations in this paper make this concrete: re-evaluating a reference in current use shows that RMSF cannot distinguish ensembles that differ substantially in state populations (Fig.~\ref{fig:rmsf}), and a model flagged as hallucinating against a short reference was vindicated once longer MD and enhanced sampling confirmed its states as physical~\cite{Diez2026}. Our claim is not that every GenMD paper exhibits these patterns, nor do we attempt to estimate their prevalence. Instead, we identify recurring evaluation practices for which the reported evidence can be insufficient to establish the corresponding scientific claim.

\paragraph{Best practices.} To improve the quality of GenMD research we lay out a sequence of best practices that ameliorate potential shortcomings of some commonly used evaluations. This includes the use of appropriate reference data for rare events, calibrating observables to slowly relaxing modes, choosing challenging benchmarks, reporting statistical measures of sampling efficiency when possible, and comparing to strong classical baselines. Finally, because training data is expensive to produce, reporting data-generation cost alongside training and inference cost is essential.

% -----------------------------------------------------------------------
\section{Background: The GenMD Landscape}
\label{sec:background}

\subsection{Molecular dynamics and the sampling problem}
MD simulations model the time evolution of molecular systems by numerical integration of equations describing the classical mechanics, yielding trajectories in phase space of positions $x$ and velocities $v$. For simplicity we restrict the discussion to positions only. The trajectories follow a time-dependent probability density $p(x_t \mid x_0)$, the Green's function of the associated Fokker--Planck equation given a fixed initial condition $x_0$. If the system evolves under a fixed potential $U$ at constant temperature, the long-time statistics converge to the Boltzmann distribution $\mu(x) \propto \exp(-\beta U(x))$, with inverse temperature $\beta$. A glossary of physics nomenclature is given in Appendix~\ref{app:glossary}.

The central purpose of MD is to complement laboratory experiments by computing averages over $\mu(x)$ and $p(x_t \mid x_0)$: \textbf{stationary observables} such as free energies and binding affinities, and \textbf{dynamical observables} such as binding rates and relaxation timescales. These averages, in principle, correspond to wet-lab measurements; the experimental observation process defines the \textit{statistic} or \textit{forward model} over which the average is computed.

Computing such high-dimensional averages typically proceeds via Monte-Carlo (MC) estimators, and this is the heart of the \emph{sampling problem}: drawing samples i.i.d.\ from $\mu(x)$ or $p(x_t \mid x_0)$ is enormously expensive. The integration time-step required for stable numerics ($\propto 10^{-15}\,\mathrm{s}$) is many orders of magnitude smaller than the experimentally relevant timescales of inter-mode mixing in the high-dimensional distribution ($t \propto 10^{-6}-10^{0}\,\mathrm{s}$). For most systems of interest, converging an MC estimator over the slow modes is impractical. Crucially, this means that for the systems GenMD aspires to model, \textit{there is no tractable ground truth against which to evaluate a generative surrogate}. This is not a technicality: it is the gap that motivates the entire field, and the reason evaluation requires special care. The gap has a natural anchor: for systems where exhaustive simulation is costly, experimental measurements are the primary available reference, the quantities a GenMD surrogate ultimately aspires to predict.

\subsection{A taxonomy of GenMD model classes}

GenMD methods are not a monolithic class; they differ in what they claim to model, and therefore in what constitutes valid evaluation. We distinguish three exemplars that bracket much of the current literature.

% \begin{description}
%   \item[Boltzmann Generators (BGs)] Learn an explicit density $\rho(x)$ approximating $\mu(x)$ and admit exact importance reweighting toward the Boltzmann density of a specified potential~\cite{noe2019boltzmann}. The quality of the approximation is measurable through the effective sample size (ESS) after reweighting.\\
%   \textbf{Goal:} make $\rho$ as close as possible to $\mu$ for a specified (generalized) potential energy model.

%   \item[Boltzmann Emulators (BEs)] Learn to \emph{sample} from $\mu(x)$ approximately, without an explicit density and without the ambition of importance reweighting~\cite{NEURIPS2022_994545b2,lewis2025scalable}. Evaluation must rely on comparing sampled statistics to independent reference data, computational or experimental.\\
%   \textbf{Goal:} recover the high-probability metastable states of $\mu$ with approximately correct Boltzmann weights. Finer details such as bond-distance and -angle statistics, and exact reweighting fidelity, are usually not the ambition.

%   \item[Implicit Transfer Operators (ITO)] Learn the transition probability $p(x_t \mid x_0)$, enabling prediction of kinetic quantities inaccessible to BGs and BEs~\cite{schreiner2023implicit,schutte2023acta,klein2023timewarp}. Meaningful evaluation requires checking dynamic observables, transients, and asymptotic behavior.\\
%   \textbf{Goal:} faithfully represent the statistics of slow dynamics on timescales $t \gg \tau$, where $\tau$ is the typical integration time of MD.
% \end{description}

\begin{methodcard}{bgcolor}{Boltzmann Generators \textnormal{(BGs)}}
Learn an \textbf{explicit density} $\rho(x)$ approximating $\mu(x)$ and admit
\textbf{exact importance reweighting} toward the Boltzmann density of a
specified potential~\cite{noe2019boltzmann}. The quality of the approximation
is measurable through the effective sample size (ESS) after reweighting.
\goal{bgcolor}{Make $\rho$ as close as possible to $\mu$ for a specified
(generalized) potential energy model.}
\end{methodcard}

\smallskip
\begin{methodcard}{becolor}{Boltzmann Emulators \textnormal{(BEs)}}
Learn to \emph{sample} from $\mu(x)$ approximately, \textbf{without an
explicit density} and \textbf{without importance reweighting}~\cite{NEURIPS2022_994545b2,lewis2025scalable,pmlr-v235-wang24cv}.
Evaluation must rely on comparing sampled statistics to independent reference
data, computational or experimental.
\goal{becolor}{Recover the high-probability metastable states of $\mu$ with
approximately correct Boltzmann weights or achieve $\mathcal{F}$-thermodynamic consistency \cite{Chennakesavalu2023,2607.19519}. Finer details---bond-distance and
-angle statistics, exact reweighting fidelity---are usually not the ambition.}
\end{methodcard}

\smallskip
\begin{methodcard}{itocolor}{Implicit Transfer Operators \textnormal{(ITO)}}
Learn the \textbf{transition probability} $p(x_t \mid x_0)$, enabling
prediction of \textbf{kinetic quantities} inaccessible to BGs and
BEs~\cite{schreiner2023implicit,schutte2023acta,klein2023timewarp,Sidky2020,Thiemann2026}.
Meaningful evaluation requires checking dynamic observables, transients, and
asymptotic behavior.
\goal{itocolor}{Faithfully represent the statistics of slow dynamics on
timescales $t \gg \tau$, where $\tau$ is the typical integration time of MD.}
\end{methodcard}

These three classes are not exhaustive but they isolate the key evaluation question: \textit{what is being sampled, and against what reference?} Methods that parameterize joint distributions over sequences of configurations (trajectory-learning approaches~\cite{jing2024generativemodelingmoleculardynamics}) sit between or beyond these classes, as do LLM-based approaches that generate conformational states without explicitly modeling thermodynamic or kinetic distributions~\cite{murtada2025mdllm}; BoPITO~\cite{diez2025bopito}, for instance, embeds a Boltzmann Generator prior within ITO learning to recover asymptotically unbiased equilibrium statistics while learning the transition kernel. In the equilibrium setting, however, the joint distribution factorizes into the stationary marginal and the one-step transition kernel, so trajectory-level similarity conflates two separable concerns: recovery of the Boltzmann marginal (a BE task) and recovery of the transition kernel (an ITO task). A principled exception arises when the model operates on a projected representation that is non-Markovian, with projected-out degrees of freedom contributing memory effects in the Mori--Zwanzig sense~\cite{zwanzig1960,mori1965}; a trajectory-level parameterization can then be motivated as a learned memory kernel.
We do not consider conformer and flexibility prediction methods~\cite{delalamo2022,stein2022speach,jing2024alphafold}, whose primary aim is mode discovery rather than calibrated probability, analogous to docking versus binding free energy calculations. When such methods claim physical fidelity, they should be held to the standards below.

An axis orthogonal to the BG/BE/ITO classification is whether a method is \emph{per-system} or \emph{transferable}. Per-system methods learn a surrogate for a single fixed molecular system; transferable methods condition on a system description and are intended to generalize across molecules, analogously to how a general force field applies to any molecule within its parameterization scope. Every class in the taxonomy above can, in principle, be either. Per-system methods are validated only on the system they were trained on whereas transferable methods must demonstrate fidelity on held-out systems to establish the generalization they claim.

%The remainder of this paper assumes only four things from this background: (i)~the true target distribution is intractable to sample exhaustively for the systems of most scientific interest; (ii)~different GenMD methods target different distributions and therefore require different evaluation; (iii)~the observables that matter scientifically (rare transitions, slow dynamics) are precisely the ones most easily missed by structural metrics; and (iv)~for systems where computational ground truth is inaccessible, experimental observables are the primary available reference, subject to the same sensitivity requirements as computational metrics.

\section{Patterns of Misaligned Evaluation}
\label{sec:problem}

We describe six recurring patterns by which current evaluation practices fail to fully support the claims practitioners intend them to support: each lets a model match what is reported while being arbitrarily wrong on what was claimed. Each is an instance of the streetlight effect (Fig.~\ref{fig:Illustration}): short trajectories stand in for converged references (P1), structural metrics for slow-mode statistics (P2), fast folders for the protein universe (P3), energy histograms for ESS (P4), unbiased MD for enhanced sampling (P5), and inference cost without transferability for total cost (P6). We deliberately do not attach individual papers to the patterns, neither as a per-paper scorecard nor as scattered example citations: the patterns are inherited community conventions, and attributing a community-level coordination outcome to individuals would be both inaccurate and counterproductive. Instead, we cite prior work that identified or analyzed a pattern, and positive exemplars of good practice. The RMSF demonstration in P2 and the per-system/transferable cost taxonomy in P6 are, to our knowledge, new; the other patterns build on prior demonstrations, such as the energy-histogram failure in P4~\cite{midgley2023flow}, which we reframe as community-level patterns. We annotate each pattern with which parts of the GenMD taxonomy it applies to (BG: \textcolor{bgcolor}{$\blacklozenge$}, BE: \textcolor{becolor}{$\bullet$}, ITO: \textcolor{itocolor}{$\blacktriangle$}  ). We then outline why this pattern is significant and provide actionable prescriptions. For convenience we summarize the prescriptions as a checklist in Appendix~\ref{app:checklist}. % Prescriptions are deferred to Section~\ref{sec:prescription}.

\subsection{P1: Distributional proxies that cannot contain the rare events being claimed \BG \BE \ITO}
\paragraph{The pattern} A widespread evaluation strategy compares model samples to a short MD trajectory rather than to representative sampling statistics. By ``short'' we mean any reference shorter than the timescales required to sample the processes a paper claims to capture. Several widely shared MD datasets fall into this category; they include MDQM9~\cite{VigueraDiez2024}, mdCATH~\cite{Mirarchi2024}, ATLAS~\cite{VanderMeersche2023}, and MISATO~\cite{Siebenmorgen2024}, with simulations on the order of $10^{1}$--$10^{2}\,\mathrm{ns}$ per system. As training corpora these are highly valuable resources. As evaluation references for claims about long-timescale behavior, however, they are blind by construction to the long-timescale processes that motivate the field: switching between metastable states, ligand binding and unbinding, folding. They are blind, in other words, to the most difficult part of MD: the rare events.

\paragraph{Why it matters} The pathology of this pattern compounds in two ways. First, if the reference does not contain a rare event, a model that correctly samples the corresponding rare state will appear to produce \emph{spurious} structures relative to the reference, penalizing exactly the behavior one wants. Recent work in this spirit~\cite{Diez2026} reports a clean illustration of the phenomenon: a model trained on short ($\sim$ns) trajectories was found to predict states absent from the training and test splits, which would conventionally be flagged as hallucinations; however, ultra-long unbiased MD and enhanced sampling subsequently confirmed those states as physical. Making the reference adequate changed the model's reported performance, here in its favor; in molecular design, this kind of re-evaluation has led to re-ranking of methods across the field~\cite{posebusters}. The closer a model comes to the true distribution, the worse it scores against a reference that missed most of it. Second, when training and evaluation use different forcefields, the comparison conflates forcefield discrepancy with modeling error: a model may be penalized for faithfully reproducing its training forcefield. Similar effects are at play when training and evaluation simulations are performed under different conditions (for example different temperatures).

\paragraph{What to do instead} A reference is fit for purpose only if it could, in principle, contain the rare events the model claims to capture. For rare-event claims, valid references include long aggregated MD on the same system; enhanced-sampling distributions (replica exchange, metadynamics, weighted ensemble~\cite{huber1996weighted}, OPES~\cite{invernizzi2020opes}) with documented convergence; experimental observables sensitive to the relevant slow motions (see P2); and large-system MSM populations and timescales. A rare-event-capable reference is therefore a question of method and of compute budget---if neither of these are available, authors should calibrate their claims accordingly. For claims restricted to local fluctuations, short MD is fit-for-purpose, but the paper should restrict its claims accordingly. Comparing generated and reference ensembles should, when possible, take the high-dimensional nature of conformational space into account \cite{lindorff2009similarity}, rather than relying on low-dimensional representations when these cannot be shown to capture most of the conformational variability. Where evaluation uses a forcefield distinct from training, the discrepancy should be reported as a separate axis, ideally by also evaluating against the training forcefield to isolate model error from forcefield mismatch. Several existing resources go beyond the short-trajectory regime and are appropriate as evaluation references on subsets of their systems: the BioEmu and CGSchNet training corpora which include extensive octapeptide datasets and much longer-trajectory subsets of CATH-derived data \cite{lewis2025scalable,Charron2025}. The CATH-derived data are not converged for all systems but may be used to build better benchmarks for assessing slow motions.

\paragraph{Precedent} These short-trajectory benchmarks have become standard partly through inheritance. An early influential benchmark established a particular evaluation protocol, and as protein-scale GenMD has matured, subsequent work has largely adopted that protocol %~\cite{jing2024alphafold,jing2024generativemodelingmoleculardynamics,lu2025structure,lu2025aligning,lu2024strstr,wolf2026learning,pmlr-v235-wang24cv,Zheng2024,feng2025biomd,Feng2026,xu2026tempo} 
rather than re-examining whether it can support the claims the field has since taken on. The pattern is a community-level coordination outcome rather than a property of any individual paper: each new method is naturally evaluated against the conventions of its predecessors, and the conventions persist by default.

\subsection{P2: Observables that are insensitive to the claimed processes \BG \BE \ITO}

\paragraph{The pattern} Diagnosing whether a model captures the relevant physics requires statistics sensitive to the slow processes that make the sampling problem hard. Two related sub-patterns recur in the literature: (i) the use of structural metrics that are mathematically insensitive to multi-basin structure, and (ii) a mismatch between the class of quantity a paper claims to model and the class of quantity its evaluation resolves.

\begin{wrapfigure}{r}{0.65\linewidth}
\vspace{-1.2cm}
    \includegraphics[width=1.0\linewidth]{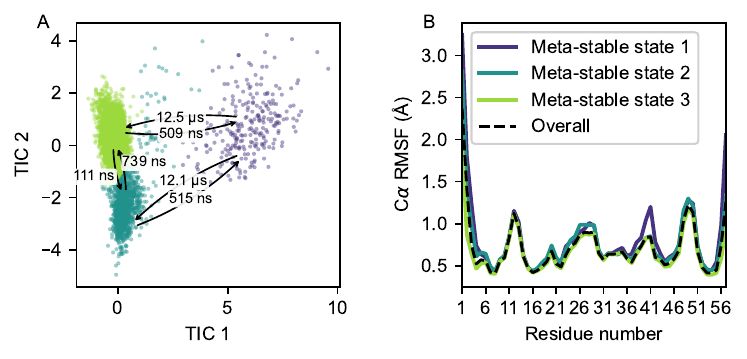}\vspace{-0.45cm}
    \caption{{\bf RMSF is a streetlight evaluator.} TICA projection of 2v0cA03 CATH domain simulation from \cite{lewis2025scalable} with three meta-stable states exchanging with mean first-passage times annotating arrows (A). Comparison of root mean square fluctuations (RMSF) within the meta-stable states (by PCCA+~\cite{Kube2007}) and for the entire ensemble (B). RMSF is nearly identical across the states and the whole ensemble, so it cannot distinguish ensembles that differ substantially in state populations or free energies: RMSF agreement is not sufficient evidence of multi-basin fidelity. \vspace{-0.7cm}}
    \label{fig:rmsf}
\end{wrapfigure}

\subparagraph{Sub-pattern A: structural metrics insensitive to multi-basin structure.} The GenMD literature commonly reports structural metrics --- root-mean-square deviation (RMSD), TM-score, and in particular root-mean-square fluctuation (RMSF) --- as primary evidence of ensemble fidelity.

\paragraph{Why it matters} RMSF is a marginal statistic of atomic positions around a mean structure: invariant to the order in which configurations are visited, with no dependence on the transition kernel $p(x_t \mid x_0)$. A model that samples a single basin with the correct local covariance, but never crosses to any other metastable state, can match the RMSF of a fully converged reference and would be indistinguishable from a correct sampler. We illustrate this point in Fig.~\ref{fig:rmsf}, using a long MD simulation reference exchanging between three different meta-stable states~\cite{lewis2025scalable}, yet the metric is largely insensitive to these changes. This observation is independent of how the samples were generated.

\paragraph{What to do instead} The principled comparison is on low-dimensional projections onto the leading eigenfunctions of the Fokker--Planck or transfer operator~\cite{schutte2023acta}. These are themselves costly, but time-lagged independent component analysis (TICA)~\cite{Perez2013} provides a data-driven linear approximation from long unbiased trajectories --- creating the obvious tension that TICA coordinates require ground-truth MD on the very systems for which GenMD surrogates are least needed. The resolution is to deploy principled metrics where available and pair them with carefully chosen physical proxies elsewhere.

For fast protein folding, the fraction of native contacts $Q(x)$ correlates strongly with the committor function~\cite{Best2013}; a threshold on $Q$ yields a reasonable folded/unfolded classifier, and a Boltzmann emulator $p(x)$ can be compared against experimental thermodynamic stability through the folding free energy $\Delta G_{\mathrm{fold}} = -k_{B}T \ln[\hat{p}_{F}/(1-\hat{p}_{F})]$, with $\hat{p}_{F} = \mathbb{E}_{x \sim p}[\mathbf{1}(Q(x) > 0.5)]$, estimated by Monte Carlo over generated samples. This is a concrete, falsifiable companion to a quantity measured in the lab, available for any protein with a solved structure. Analogous proxies exist for binding (interface contacts, umbrella-sampling order parameters) and cryptic pocket opening (pocket volume along a reference coordinate). More generally, statistics sensitive to multi-basin structure --- metastable state populations, free-energy differences between identified states, autocorrelation functions along slow collective coordinates, fluxes between states --- resolve precisely the failure mode that RMSF cannot~\cite{lindorff2009similarity,tiberti2015encore}. Where structural metrics are reported, they should be accompanied by at least one such multi-basin statistic.

\subparagraph{Sub-pattern B: thermodynamic evaluation for kinetic claims. \ITO} Equilibrium densities and transition kernels are different mathematical objects, and agreement on the former implies little about the latter: a model can reproduce $\mu(x)$ exactly while getting every rate and relaxation timescale wrong. Boltzmann-Emulator-class models in the recent literature often validate against thermodynamic observables while making claims about protein {\it dynamics}.

\paragraph{Why it matters} A model whose stationary distribution is correct can still place all dynamical weight on the wrong relaxation pathways, predict barriers that are wrong by orders of magnitude, or invert the ordering of slow timescales --- none of which is detectable from $\mu(x)$ alone. Reporting thermodynamic agreement remains necessary, and propagator invariance of the Boltzmann distribution is a meaningful sanity check, but these tests leave the kinetic content of the model untested.

\paragraph{What to do instead} The natural kinetic proxies are autocorrelation functions of slow collective coordinates, mean first-passage times between metastable basins, and relaxation timescales --- as reported by small-system surrogates~\cite{schreiner2023implicit,Diez2026,klein2023timewarp} or via Markov state model fluxes in peptide-scale evaluations~\cite{jing2024generativemodelingmoleculardynamics}. As models scale to proteins, this style of evaluation should scale with them: reporting at least one kinetic observable is the difference between a thermodynamic and a kinetic claim.

The same sensitivity argument applies to experimental observables. Ensemble- and time-averaged measurements (NMR chemical shifts, SAXS profiles, residual dipolar couplings) constrain the stationary distribution and are appropriate for thermodynamic claims, but are blind to the transition kernel: kinetically distinct models can produce identical averaged signals. Kinetics-sensitive experiments (NMR relaxation~\cite{palmer2004nmr}, single-molecule fluorescence correlation spectroscopy~\cite{schuler2013fret}, time-resolved techniques~\cite{gusew2026integrative}) resolve the slow modes and are the appropriate substitute when no computational kinetic reference is available, provided a forward model connects simulation configurations to the measured quantity. For example, FRET efficiency predicted from generated configurations via a structural dye-pair model can be directly compared to single-molecule measurements to test whether a model captures the correct conformational interconversion timescales~\cite{schuler2013fret,gusew2026integrative}.

\paragraph{Precedent} The pattern is reinforced by the asymmetric difficulty of the underlying calculations: structural metrics are cheap to compute from any sample of configurations, while kinetic observables typically require a Markov state model built from the generated samples or knowledge of a slowly relaxing collective variable. As with P1, the convention persists in part because the easier evaluation is the one that travels between papers.

%\paragraph{Transition path validation.} A parallel pattern affects transition-path methods. Several approaches~\cite{holdijk2023stochastic,du2024doobs,raja2025actionminimization} guide generative models through Lagrangian-based ans\"atze to produce pathways connecting metastable states. These methods typically demonstrate qualitative plausibility (generated paths visit physically reasonable intermediates and reach the target state) but quantitative validation against kinetic observables is rare. A path that traverses a plausible intermediate while implying the wrong barrier height, and therefore the wrong rate by orders of magnitude, is not a useful surrogate for dynamics. Even more common are methods that do not provide any information on the free energy along a path. The gap is sharpest where the theoretical foundation is sound but empirical validation does not reach the quantities the theory targets: an approximation to Doob's $h$-transform~\cite{du2024doobs} would, if accurate, imply the correct rate, yet evaluations in this literature commonly report path plausibility and energy profiles rather than recovered rates or barrier heights. The classical transition path sampling and forward flux sampling literature provides validated rate estimates for many of the same systems used as ML demonstrations~\cite{vanErp2003,Allen2009}; comparison against these references is the natural way to close the loop. Further, the field need improved methods for comparing trajectories on complex landscapes \cite{onuchic1999analyzing}.

\subsection{P3: System selection narrower than the scope of the claims \BG \BE \ITO}

\paragraph{The pattern} Papers overwhelmingly benchmark on small, fast-relaxing systems (fast-folding proteins~\cite{lindorff2011fast} and tetra-peptide datasets~\cite{klein2023timewarp}) where MD is costly but in principle a ground truth is available. These are excellent benchmarks, especially when reserved for evaluation, and a necessary stepping stone toward better GenMD methods. They are also a limited and unrepresentative slice of molecular systems, and claims made on the basis of these benchmarks should reflect that. The pattern we want to flag is not the use of small benchmarks: it is the inferential leap from benchmarked performance on small systems to claims about general protein-scale capability. Benchmarking on small fast-relaxing systems is necessary, yet success does not imply general protein-scale capability.

\paragraph{What to do instead} The first remedy is to state the system scope explicitly and confine claims to it. Where larger-scale claims are made, hybrid evaluation is the natural path: principled metrics on small systems for fidelity verification (where ground truth is accessible), paired with physical proxies (P2) and experimental observables sensitive to the relevant timescales on representative larger systems. Where transferability is claimed, at least one held-out evaluation system should differ from the training distribution along an axis the method claims to generalize over (sequence length, fold class, ligand chemistry) rather than only across random splits within the training distribution. Reporting performance separately on in-distribution and out-of-distribution systems clarifies which aspects of the claim the evaluation actually supports.

\subsection{P4: Avoiding the one reference-free metric available \BG}

\paragraph{The pattern} The status of the effective sample size as a reference-free metric primarily concerns Boltzmann Generators. The target distribution is specified by a known potential $U(x)$ (a force field), so the target Boltzmann density $p(x) \propto e^{-U(x)/k_\mathrm{B} T}$ is available without any reference trajectory. The self-normalized (reverse) ESS is directly computable and directly gauges the {\it local} quality of a BG. This is the one reference-free diagnostic available when the true baseline is costly, and it is the gold-standard metric for the BG class. It is not, however, self-correcting: estimated from model samples, ESS only sees regions the model visits, so a flow that has dropped a metastable state can still report a high ESS~\cite{midgley2023flow}.

A separate but related issue is that several recent methods fit the BG definition --- learning an explicit density that targets a specified Boltzmann distribution and admits importance reweighting --- without self-identifying as Boltzmann Generators in their titles or framing. We include them under the BG umbrella because the evaluation criteria that apply to BGs apply to them: ESS reporting, reweighting fidelity, and overlap with the target density.

\paragraph{Why it matters} The literature has bifurcated. BGs admit exact likelihood evaluation, making ESS reporting an obligation that the best current work meets~\cite{midgley2023flow,TransferableBoltzmannGen,tan2025scalable,schopmans2025temperatureannealed,klitzing2026learning,schopmans2026efficient}. What is not constructive --- and is the pattern we flag here --- is substituting energy histogram comparisons as though they provide equivalent assurance to ESS in cases where ESS is the appropriate diagnostic. By concentration of measure, a model can produce samples in the correct energy range while placing them in entirely wrong regions of configuration space, and mode collapse to the dominant basin may be invisible to energy histograms. A related failure has been demonstrated explicitly on small systems~\cite{midgley2023flow}, where a flow trained by maximum likelihood obtained good test log-likelihood and Ramachandran-plot agreement while mis-modeling a low-population metastable state, yet had an ESS of only ${\approx}3\%$. Where energy histograms appear without complementary validation in protein-scale evaluations, %~\cite{pmlr-v235-wang24cv,Zheng2024}
the metric cannot detect the failure mode it is implicitly being used to rule out.

\paragraph{What to do instead} For any architecture admitting log-likelihood evaluation against a known target density, the reverse ESS is the diagnostic of choice and should be reported, alongside a reference comparison (P2) wherever one exists, since ESS alone cannot detect missing modes. For BEs the weakness should be acknowledged explicitly~\cite{lewis2025scalable} and compensated for by reference-comparison evaluations against multi-basin statistics (P2), not by energy histograms alone. For ITOs, ESS in the BG sense is not the right metric, and class-appropriate diagnostics (P2) take its place.

\subsection{P5: Absent or weak classical baselines \BG \BE \ITO}
\paragraph{The pattern} GenMD methods are routinely compared to brute-force MD or to other generative models, but rarely to the classical enhanced-sampling and analysis methods that represent the practical state of the art for the same observables. Markov state models~\cite{prinz2011markov, hoffmann2021deeptime}, replica exchange MD~\cite{Sugita2019}, metadynamics and its descendants~\cite{Laio2002,Grubmller1995,invernizzi2020opes}, transition interface sampling~\cite{vanErp2003}, and forward flux sampling~\cite{Allen2009} have decades of methodological development and well-understood failure modes. They are the methods practitioners would otherwise use to compute the quantities GenMD now claims to accelerate; they are the relevant baseline.
 
\paragraph{Why it matters} The omission of classical baselines makes it impossible to assess whether a new method provides genuine acceleration over the practical state of the art or merely repackages existing capability in a more expensive framework. A Boltzmann emulator that estimates a folding free energy faster than $\mu\mathrm{s}$ of unbiased MD but worse than well-calibrated metadynamics or replica exchange in $100\mathrm{s}$ of $\mathrm{ns}$ does not provide a meaningful advantage: the apparent speedup over MD is real but practically irrelevant, because the realistic alternative the practitioner faces is the classical method, not unbiased MD. Reporting only the easier comparison conflates the question ``does this method beat MD?'' with the question that actually matters, ``should this method be run instead of an enhanced-sampling protocol the field already has?''.
 
\paragraph{What to do instead} At least one classical enhanced-sampling baseline should be included for any task on which a generative method claims acceleration. The choice of baseline should target the same observable as the generative method --- replica exchange or metadynamics for thermodynamic quantities, MSM-driven adaptive sampling for kinetic spectra, transition interface or forward flux sampling for rate constants --- and should use a setup with documented convergence on the system class in question. The comparison should be made at matched compute or matched observable quality, stating which; otherwise it reduces to a difference in tuning effort. Where a single matched budget is not decisive, an accuracy-vs-budget curve should be reported. This fairness budget constrains the methods being compared, the GenMD method and its classical baseline, both scored against the same fixed and validated reference: a reference is produced once, offline, at a published cost, and thereafter cited rather than re-run (Section~\ref{sec:benchmarks}). Where the classical method has tuning hyperparameters (collective variables for metadynamics, replica spacing for REMD), the choice should be reported and ideally reflect community-standard configurations rather than ad hoc settings, so that the comparison reflects the practical state of the art.
 
\paragraph{Precedent} This pattern is among the most pervasive: it is rare for a GenMD paper to compare against even one enhanced-sampling method, and rarer still for the comparison to control for compute. The root cause is partly the absence of shared benchmarks specifying both the system and the  budget.

\subsection{P6: Cost accounting that does not reflect how the method will be used \BG \BE \ITO}

\paragraph{The pattern} GenMD papers routinely report inference-time speedups (``$1000\times$ faster than MD'') without accounting for training data generation, training compute, or the conditions under which the inference speedup translates into a real saving. The correct accounting depends on whether the method is per-system or transferable, and the current literature largely fails to make this distinction.

For \emph{per-system methods} (most current BGs, and per-system BEs and ITOs), training cost is paid once per system and amortized across queries on that system. The relevant currencies are data-generation cost $C_{\text{data}}$ (MD time for the training set), training cost $C_{\text{train}}$ (GPU-hours), and per-query inference cost $C_{\text{inf}}$. The break-even query count $N_{\text{break}} = (C_{\text{data}} + C_{\text{train}})/S_{\text{query}}$, with $S_{\text{query}} = C_{\text{inf}}^{\text{ref}} - C_{\text{inf}}^{\text{ML}}$, is well-defined and informative.

For \emph{transferable methods} (most BEs, transferable BGs and ITOs), the accounting changes in kind. Data generation cost is structurally analogous to force-field parameterization: a one-time investment reused across an open-ended user base. We do not compute amortization horizons for AMBER~\cite{cornell1995amber} or CHARMM~\cite{mackerell1998charmm}, not because the cost is small but because the user base is not enumerable in advance. The same applies to transferable GenMD models, and a per-system $N_{\text{break}}$ is not the right object; the relevant question is whether inference is competitive with classical enhanced sampling at matched compute or matched observable quality.

\paragraph{Why it matters} For per-system methods, reporting only $C_{\text{inf}}$ obscures whether the upfront cost is ever recouped: a $1000\times$ inference speedup on a system requiring only a handful of queries can leave the method net more expensive than the baseline. For transferable methods, the force-field analogy holds only when the method actually generalizes: a model demonstrated only on systems closely resembling its training distribution has not established the breadth that makes the prepaid framing appropriate, and $C_{\text{data}}$ should be charged against the limited scope of demonstrated generalization. There is a spectrum of thresholds: $C_{\mathrm{data}}$ is amortized over the number of systems on which out-of-distribution fidelity is demonstrated within errors stated in advance, ranked by difficulty (random split within the training distribution $<$ held-out fold or scaffold $<$ unseen size regime $\leq$ unseen chemistry). Conditioning on pre-trained representations (protein language model embeddings~\cite{antoniadis2026platito} or aligned force--conformation representations~\cite{pinede2025unifying}) is a promising direction for extending generalization at low additional cost; establishing when and how much such representations help is an open empirical question, and out-of-distribution evaluation is the natural tool for answering it.

\paragraph{What to do instead} For per-system methods: report $C_{\text{data}}$, $C_{\text{train}}$, and $C_{\text{inf}}$ explicitly, state the reference method against which $S_{\text{query}}$ is computed, and report $N_{\text{break}}$. For transferable methods: report $C_{\text{data}}$ and $C_{\text{train}}$ as transparent upfront investments; compare against MD and at least one classical enhanced-sampling method (P5) \emph{at matched compute or matched observable quality}; and demonstrate the transferability that justifies the prepaid framing through out-of-distribution evaluation (P3).

\section{The Benchmark Gap}
\label{sec:benchmarks}

The most consequential gap is the absence of curated benchmark suites spanning the application scope GenMD claims to address. Without shared yardsticks, the patterns of Section~\ref{sec:problem} are the natural equilibrium: each method invents its own evaluation, and short-trajectory references travel unexamined between papers.

A useful suite is a curated collection of challenging systems, modular by application scope: (i)~\textbf{proteins} with well-characterized slow conformational change (domain motions, cryptic pocket opening, allosteric switching, slow folding), of which fast-folding two-state proteins~\cite{lindorff2011fast} are a tractable subset, not the whole suite; (ii)~\textbf{protein--ligand systems} with measured binding affinities and residence times; (iii)~\textbf{materials systems} with slow collective rearrangements (solid--solid transitions, defect migration, nucleation)---heavy MD users, largely absent from current GenMD benchmarks; and (iv)~\textbf{further domains} (membranes, polymers, glasses, liquids) as the field expands.

Each system must specify at least one observable of scientific interest (folding free energy, transition rate, phase boundary), together with a rigorously evaluated reference value and its uncertainty.

\emph{Computational references} must publish the total compute cost incurred to obtain the value, alongside the protocol used (force field, sampling method, hardware, convergence diagnostics). That published cost is the budget a generative method must beat, at a matched observable quality, to claim acceleration, tying benchmarks directly to the cost accounting of P6.

\emph{Experimental references} are appropriate, and often the only option, where computational ground truth is costly, with the analogous requirement: measurement protocol, forward model, and uncertainty. Ensemble averages (e.g. chemical shifts) validate thermodynamic claims; kinetics-sensitive measurements~\cite{palmer2004nmr,schuler2013fret,feng2026cooperative} are required for kinetic ones. PeptoneBench~\cite{invernizzi2025peptonebench} is a useful template.

A suite built this way addresses Section~\ref{sec:problem}: references can contain the rare events being claimed (P1), observables resolve the slow processes (P2), the system set spans the claimed scope (P3), classical reference values are explicit (P5), and reference compute sits alongside ML compute in any acceleration claim (P6). Released in versioned form, such references can be cited rather than rerun.

The benchmark need not be perfect to be useful; it need only specify the question well enough that papers can be compared. An institutional model already exists in this domain: community efforts to make biomolecular simulation data FAIR~\cite{amaro2025need} and to find and index simulation data already deposited in public repositories~\cite{tiemann2024mdverse,MDposit2025}. The unit of contribution is a reference record rather than a paper: system, observable, value with uncertainty, simulation and analysis protocol, convergence diagnostics, and total compute. Most records can be curated from existing data rather than simulated afresh, and versioned hosting is already part of the checklist (Appendix~\ref{app:checklist}, Data and protocol availability).

\section{Alternative Views}
\label{sec:alternatives}
\paragraph{``It is too early; rigor now will slow the field.''} GenMD is young, and demanding long references, classical baselines, and full cost accounting from every paper raises the cost of exploratory work. \emph{Concession:} some checklist items are costly, and small, fast-relaxing benchmarks remain a necessary stepping stone toward better methods (P3). \emph{Reply:} the checklist is tiered (Appendix~\ref{app:checklist}): Tier~A costs almost nothing, and costly items are required only when the matching claim is made. Once a tractable evaluation exists, it resists questioning, because questioning it invalidates a chain of comparable results; the time to set standards is before today's conventions calcify into norms.

\paragraph{``Without tractable ground truth, no standard is enforceable.''} For the systems GenMD aspires to model there is no tractable ground truth (Section~\ref{sec:background}), so a standard that presumes one cannot be checked. \emph{Concession:} where no fit-for-purpose reference exists, no standard can validate a rare-event claim; the claim must shrink accordingly. \emph{Reply:} a rare-event-capable reference is a question of method, not of raw compute (P1), and it need not be globally converged: the data in Fig.~\ref{fig:rmsf} are not, yet they resolve three slowly interconverting metastable states, so a model that reproduces their populations and interconversion timescales can claim to capture that system's slow dynamics on that timescale. Some checks need no reference at all: ESS can be computed for proteins where no long-timescale MD has ever been run (P4). Enforcement is adoption at review time: reviewers and area chairs can require Tier~B and~C items when the corresponding claim is made.

\paragraph{``Short-trajectory benchmarks are adequate for the applications that matter.''} Many applications need only local fluctuations or the dominant state, which short trajectories sample well. \emph{Concession:} short-trajectory datasets are highly valuable training corpora, and for claims restricted to local fluctuations short MD is fit for purpose (P1). \emph{Reply:} such papers should then restrict their claims accordingly, and evaluate whether they provide tangible value beyond classical references.

%\paragraph{``Benchmarks fossilize fields.''} A measure originally chosen for convenience becomes a target, and optimization pressure drifts the field toward what the measure rewards. \emph{Concession:} the risk is real, and it applies to the benchmarks we propose. \emph{Reply:} the same drift is already under way with short-trajectory references. The suite we propose is modular by application scope and built from versioned reference records that can be extended and revised (Section~\ref{sec:benchmarks}); it need not be perfect to be useful, only specify the question well enough that papers can be compared.

\section{Conclusion}
The GenMD community is not properly evaluating its progress: current evaluations let a model match what is reported while being arbitrarily wrong on what was claimed. The misalignment is structural rather than incidental, but fixable. The patterns we have described (short references, slow-mode-blind metrics, narrow system selection, avoidance of self-correcting metrics, weak classical baselines, and unreported costs) are not properties of individual papers but of the equilibrium produced by the field's current incentives. Naming the patterns is the first step; specifying observables, baselines, and cost accounting is the second; building shared benchmark infrastructure is the third. That infrastructure has a concrete shape: curated suites spanning the field's application scope, with reference values rigorously evaluated and, where computational, the cost of obtaining them transparently reported.

Evaluation norms are still being established, and the choices made now will shape what counts as progress for years to come. We call on the community to use the tiered checklist proposed here at review time, to encourage classical baselines as a matter of course, and to invest in shared benchmarks, both computational and experimental, that make the hard cases visible. The science is too important to evaluate where the light is good rather than where the keys are.

\begin{ack}
    SO thanks C. von Klitzing for comments on a late version of this manuscript and acknowledges funding from the European Union (ERC, SPETITO, 101230596), funding from the Wallenberg AI, Autonomous Systems and Software Program (WASP) funded by the Knut and Alice Wallenberg Foundation, and the Chalmers Academic Excellence Initiative. KL-L acknowledge support by the European Union (ERC, DynaPLIX, SyG-2022 101071843) and the Novo Nordisk Foundation for the PRISM centre (NNF18OC0033950). Views and opinions expressed are however those of the author(s) only and do not necessarily reflect those of the European Union or the European Research Council. Neither the European Union nor the granting authority can be held responsible for them. This research was
supported by the National Institute of General Medical Sciences of the NIH under award number 1R35GM159834-01 (GMR). The content is solely the responsibility of the authors and does not necessarily represent the official views of the NIH.
\end{ack}

\printbibliography
\clearpage
\appendix

\section{A tiered evaluation checklist}
\label{app:checklist}
The checklist has three tiers. Tier~A is always required and costs almost nothing; Tier~B is required when the matching claim is made; Tier~C is encouraged, with an escape hatch. We do not rely on the checklist changing behavior on its own: Tier~A is cheap enough to serve as internal review criteria, and reviewers and area chairs can require Tier~B and~C items when the corresponding claim is made.

% --- Body ---
\begin{tcolorbox}[
  enhanced,
  colback=checkcolor!2!white,
  colframe=checkcolor!55,
  coltitle=white,
  colbacktitle=checkcolor,
  title={\textbf{Evaluation checklist}},
  fonttitle=\small\sffamily,
  boxrule=0.5pt, arc=2pt,
  left=10pt, right=10pt, top=8pt, bottom=10pt,
  attach boxed title to top left={xshift=8pt, yshift=-\tcboxedtitleheight/2},
  boxed title style={arc=2pt, boxrule=0pt},
]

\checksection{Tier A: always}
\begin{checklist}
  \item \chkitem{Target distribution}{State formally what the model samples from.}
        {\small What distribution does the model claim to sample from---$\mu(x)$,
        $p(x_t \mid x_0)$, both, or something else?}

  \item \chkitem{System scope}{Justify that the benchmarks cover the claimed scope.}
        {\small If the model is proposed for general proteins, why is it
        evaluated only on fast-folding proteins?}

  \item \chkitem{In-distribution vs.\ out-of-distribution}{Distinguish ID from
        OOD claims explicitly.}
        {\small A model trained on small peptides and tested on small peptides
        has not demonstrated generalization to proteins.}

  \item \chkitem{Cost accounting}{Report $C_{\text{data}}$, $C_{\text{train}}$,
        and $C_{\text{inf}}$ separately.}
        {\small For per-system methods: name the classical reference method and
        state $N_{\text{break}}$. For transferable methods: amortize $C_{\text{data}}$
        over the systems on which out-of-distribution fidelity is demonstrated
        within errors stated in advance, and state the most difficult split
        reached (random split $<$ held-out fold or scaffold $<$ unseen size
        regime $\leq$ unseen chemistry).}

  \item \chkitem{Data and protocol availability}{Release datasets, code, and
        evaluation protocols on standard platforms.}
        {\small Deposit on, e.g., HuggingFace or Zenodo, along with predicted
        summary statistics (free energies, relaxation timescales, observable
        averages) in a reusable, versioned form.}
\end{checklist}

\checksection{Tier B: when the claim is made}
\begin{checklist}
  \item \chkitem{Reference fitness}{Check that the reference data could in
        principle contain the rare events or slow processes being claimed.}
        {\small A reference shorter than the timescales needed to sample the
        claimed processes is blind to them by construction; short MD is
        fit for purpose only when claims are explicitly restricted to local
        fluctuations.}

  \item \chkitem{Class--metric alignment}{Use thermodynamic metrics for
        thermodynamic claims, kinetic metrics for kinetic claims.}
        {\small Report at least one observable in each category being claimed.
        Where structural metrics are reported, accompany them with at least one
        multi-basin statistic (state populations, free-energy differences, or
        autocorrelation functions along slow collective coordinates).}

  \item \chkitem{ESS where possible}{For BGs 
        report ESS after reweighting.}
        {\small A model with high visual fidelity but ESS $\ll N$ is not a
        functioning importance sampler.}
\end{checklist}

\checksection{Tier C: encouraged}
\begin{checklist}
  \item \chkitem{Classical baselines}{Compare to a strong classical method at
        matched compute or matched observable quality, stating which.}
        {\small At least one of: Markov state models, replica exchange,
        metadynamics, transition interface sampling. Where a single matched budget
        is not decisive, report an accuracy-vs-budget curve. If no baseline can be
        run, state which classical method a practitioner would use and why it was
        not run.}

  \item \chkitem{Calibrated conclusions}{Claim--evidence alignment.}
        {\small  Restrict conclusions to the processes, timescales, systems, and observables that the evaluation can resolve.}
\end{checklist}

\end{tcolorbox}

\section{Glossary}
\label{app:glossary}

\begin{tcolorbox}[
  enhanced,
  colback=glosscolor!3!white,
  colframe=glosscolor!55,
  coltitle=white,
  colbacktitle=glosscolor,
  title={\textbf{Glossary of physics nomenclature}},
  fonttitle=\small\sffamily,
  fontupper=\small,
  boxrule=0.5pt, arc=2pt,
  left=8pt, right=8pt, top=6pt, bottom=8pt,
  attach boxed title to top left={xshift=8pt, yshift=-\tcboxedtitleheight/2},
  boxed title style={arc=2pt, boxrule=0pt},
]

\glosscat{State space}
\begin{description}[style=unboxed, leftmargin=0pt, labelindent=0pt,
                    itemsep=2pt, parsep=0pt, labelsep=0.5em]
  \item[{\term{Phase space}}]
    All possible states of the system, parameterised here by atomic positions
    $x$ (and, in the full treatment, velocities $v$).
  \item[{\term{Metastable state}}]
    A contiguous region of phase space in which the system is stable for a
    long time relative to the fastest time scales: a valley in the landscape.
\end{description}

\glosscat{Dynamics \& kinetics}
\begin{description}[style=unboxed, leftmargin=0pt, labelindent=0pt,
                    itemsep=2pt, parsep=0pt, labelsep=0.5em]
  \item[{\term{Langevin equation}}]
    A stochastic differential equation describing a system evolving under a
    deterministic force $F(x) = -\nabla U(x)$ plus thermal noise at
    temperature $\beta^{-1}$.
  %\item[{\term{Trajectory}}]
   % A sampled path $(x_0, x_{\Delta t}, _{2\Delta t}, \ldots)$ obtained by
   % numerical integration of the Langevin equation from initial condition $x_0$.
  \item[{\term{$p(x_t \mid x_0)$}}]
    The \emph{transition density}: probability of observing state $x_t$ at
    time $t$ given initial state $x_0$. Encodes the system's \emph{kinetics}.
  \item[{\term{Fokker--Planck equation}}]
    The deterministic PDE governing the evolution of $p(x_t \mid x_0)$. Its
    fundamental solution is the \emph{Green's function} of the system.
\end{description}

\glosscat{Thermodynamics}
\begin{description}[style=unboxed, leftmargin=0pt, labelindent=0pt,
                    itemsep=2pt, parsep=0pt, labelsep=0.5em]
  \item[{\term{Boltzmann distribution}}]
    The stationary distribution $\mu(x) \propto \exp(-\beta U(x))$ reached as
    $t \to \infty$. Encodes the system's \emph{thermodynamics}.
  \item[{\term{$\beta$}}]
    Inverse temperature, $\beta = 1/(k_\mathrm{B} T)$.
    Higher $\beta$ $=$ colder system $=$ sharper Boltzmann distribution.
\end{description}

\glosscat{Observation}
\begin{description}[style=unboxed, leftmargin=0pt, labelindent=0pt,
                    itemsep=2pt, parsep=0pt, labelsep=0.5em]
  \item[{\term{Forward model}}]
    A model that predicts a microscopic instantaneous observable from a
    molecular configuration.
  \item[{\term{Observable}}]
    A function of configuration whose expectation under $\mu$ (stationary)
    or $p(x_t \mid x_0)$ (dynamic) corresponds to an experimentally
    measurable quantity. Some observables are time-resolved, with a
    trajectory observed through a forward model.
\end{description}

\end{tcolorbox}

\end{document}